\documentclass[fleqn,twoside]{article}
\usepackage{espcrc2}
\usepackage{psfig}

\usepackage{graphicx}
\usepackage[figuresright]{rotating}

\def\xmm{{\em XMM-Newton}~}
\def\xte{{\em RXTE}~}
\def\mch{M$\rm^{c}$Hardy\,}
\def\etal{\rm et al.~\rm}

\newcommand{\AmS}{{\protect\the\textfont2
  A\kern-.1667em\lower.5ex\hbox{M}\kern-.125emS}}

\title{Combined Long and Short Timescale X-ray Variability of NGC~4051 with
{\it RXTE} and {\em XMM-Newton} }

\author{I.M. M$\rm^{c}$Hardy\address[soton]{Department of Physics 
and Astronomy, The University, Southampton SO17 1BJ},
       I.E. Papadakis\address{Physics Department, University of Crete,
       Heraklion, Crete, Greece},
       P. Uttley\addressmark[soton],
       K.O. Mason\address[mssl]{Mullard Space Science Laboratory, 
University College London, Holmbury St Mary, Dorking RH5 6NT},
        and M.J. Page\addressmark[mssl]}
       
\begin{document}

\begin{abstract}
We present a comprehensive examination of the X-ray variability of the
narrow line Seyfert 1 (NLS1) galaxy NGC~4051. We combine over 6.5
years of frequent monitoring observations by {\it RXTE} with a
$>100$~ks continuous observation by {\em XMM-Newton} and so present a
powerspectral density (PSD) covering an unprecedent
frequency range of over 6.5 decades from $<10^{-8}$ to $>10^{-2}$ Hz.
The combined {\it RXTE} and {\em XMM-Newton} PSD is a very good match
to the PSD of the galactic black hole binary system (GBH) Cyg~X-1 when
in a `high', rather than `low', state providing the first definite
confirmation of an AGN in a `high' state.

We find a break in the PSD at a frequency
$\nu_{B}=8^{+4}_{-3}\times10^{-4}$Hz.  If $\nu_{B}$ scales linearly
with mass then, assuming a black hole (BH) mass of $10 M_{\odot}$ for Cyg X-1,
we imply a BH mass of $3^{+2}_{-1} \times 10^{5}
M_{\odot}$ in NGC~4051, which is consistent with the recently reported
reverberation value of $5^{+6}_{-3} \times 10^{5} M_{\odot}$.
Hence NGC~4051 is emitting at $\sim30\% \rm~ L_{Edd}$.

We note that the higher energy photons lag the lower energy ones and
that the lag is greater for variations of longer Fourier period and
increases with the energy separation of the bands.  Variations in
different wavebands are very coherent at long Fourier periods but the
coherence decreases at shorter periods and as the energy separation
between bands increases.  This behaviour is similar to that of
GBHs and suggests a radial distribution of frequencies and photon
energies with higher energies and higher frequencies being associated
with smaller radii.

It is not possible to fit all AGN to the same linear scaling of break
timescale with black hole mass.  Broad line AGN are consistent with a
linear scaling of break timescale with mass from Cyg~X-1 in its low
state but NLS1 galaxies scale better with Cyg~X-1 in its high
state. We suggest that the relationship between black hole mass and
break timescale is a function of another underlying parameter which
may be accretion rate.
\vspace{1pc}
\end{abstract}

\maketitle

\section{INTRODUCTION}
 
Early observations with EXOSAT, coupled with some archival data,
indicated a similarity between the powerspectral densities (PSDs) of
AGN and galactic black hole X-ray binary systems (GBHs). Both types of
PSD were steep at high frequencies and flattened, below a break or
`knee' frequency \cite{mch88}.
Early indications were that the break
timescale scaled with black hole mass but the long timescale (weeks -
years) X-ray lightcurves of AGN were not of adequate quality to
properly determine the break timescales. However since the launch of
the Rossi X-ray Timing Explorer ({\it RXTE}; \cite{swank98})
in December
1995 we have been able to obtain excellent long timescale AGN
lightcurves, and PSDs, and evidence is building for a correlation between black
hole mass and break timescale (eg \cite{mch98,edelson99,mark03}).

GBHs, however, are found in at least two states: the low flux, hard
spectrum (`low') state and the high flux, soft spectrum (`high') state
(eg \cite{mcclintock03,nowak99,cui97b}).  The PSDs of these two states
are different. At high frequencies both PSDs have slopes of $\sim
-2.5$. Below a break frequency ($\nu_{B}\sim3$Hz for the low state and
$\sim15$Hz for the high state) the PSDs both flatten to a slope of
-1. In the high state there is no other observable slope change but in
the low state the PSD flattens further (below $\nu \sim 0.3$Hz) to a
slope of zero.  It had been thought, based on the similarity of their
energy spectra, that AGN were the analogues of GBHs in their low
state. However, by combining \xte and \xmm observations, we show that 
NGC4051 is the analogue of a high state GBH. The PSD
we present of NGC4051 is the best PSD yet produced for an AGN. In this
paper we consider the implications of our result for the understanding
of the similarity between AGN and GBHs. We also examine the variation,
with energy, of the high frequency PSD slope and the coherence of the
X-ray emission, and draw conclusions regarding the geometry of the
emitting region and the origin of theblack holemass/break timescale relationship.

\section{OBSERVATIONS}
\label{sec:obs}

\begin{figure}
\psfig{figure=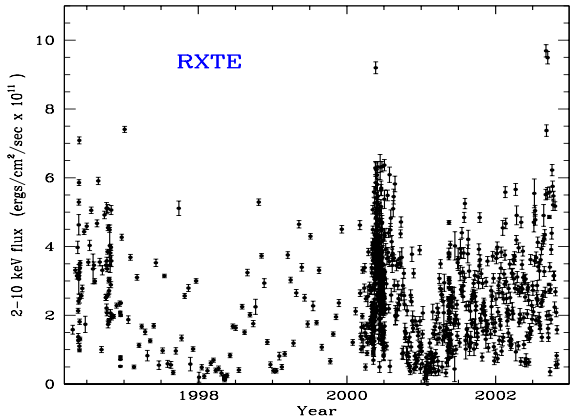,width=8cm,height=6cm,angle=0}
\caption{\footnotesize {\it RXTE} Long Term 2-10 keV lightcurve of NGC~4051. 
Each data point represents an observation of $\sim1$~ks.}
\label{fig:4051xtelc}
\end{figure}

\begin{figure}
\psfig{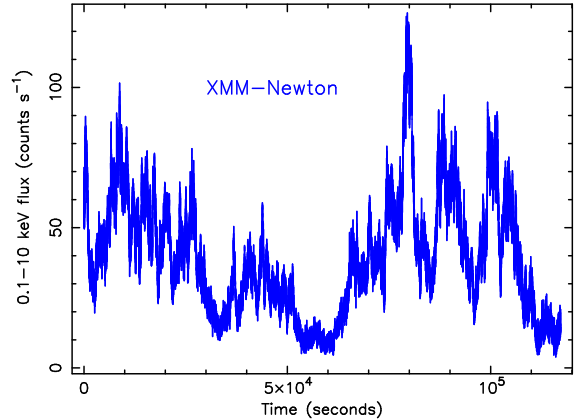}
\caption{\footnotesize XMM-Newton background-subtracted lightcurve 
of NGC~4051 in the 0.1-10 keV energy band, with 5s time bins.}
\label{fig:4051xmmlc}
\end{figure}

In Fig.~\ref{fig:4051xtelc} we present all of the short
($\sim$1~ks duration) monitoring observations of NGC~4051 which have
been carried out with the proportional counter array (PCA, 
\cite{zhang93})
on {\it RXTE} from 1996 until 2002 inclusive.
In Fig.~\ref{fig:4051xmmlc} we present the full-band \xmm lightcurve
obtained from a continuous observation over the 16th and
17th May 2001. The lightcurve is the sum of data from the PN and MOS
detectors. We refer readers to \cite{mch03} for details of the
analysis of the \xte and \xmm data.

\section{POWERSPECTRAL ANALYSIS}

\subsection{Combined \xte and \xmm PSD}

Powerspectral analysis of the continuous \xmm lightcurves is straighforward
but analysis of the \xte lightcurves is more difficult as they consist
of many short observations with a variety of sampling patterns in
order to cover different Fourier periods.  Since early 2000 we have
had observations approximately once every 2 days, and we have had a
period (mid 2000) of 2 months where we observed every 6 hours. However prior to
2000 the observations were less frequent.
To cope with the distortions which the sampling pattern
introduces into the PSD, we employ a simulation-based modelling
technique
\cite{uttley02}.  We are able to incorporate the \xmm observations
into our modelling. We again refer to \cite{mch03} for details.
Below (Section~\ref{sec:xmmpsd}) we show
that $\nu_{B}$ is independent of energy
and so here we use the 0.1-2 keV \xmm data which shows the break most clearly.

\begin{figure}
\psfig{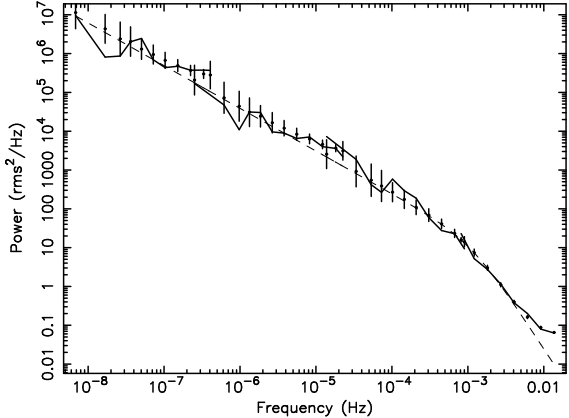}
\caption{\footnotesize Combined {\it RXTE} and {\em XMM-Newton} 
PSD. The underlying, undistorted, model is given by the
smooth dashed line. The observed data is given by the continuous jerky
line and the model, distorted by the sampling pattern, is given by the
points with errorbars. }
\label{fig:xtexmm012psd}
\end{figure}

A simple powerlaw, although a reasonable fit to the low frequency \xte
PSD on its own, is a very poor fit to the combined \xte and \xmm
PSD. The best fit (Fig~\ref{fig:xtexmm012psd}) is to a bending
powerlaw \cite{mch03} where the PSD changes smoothly from a slope of
$\alpha_{L}=-1.1$ below, to a slope of $\alpha_{H}=-2.8$ above
$\nu_{B}=8^{+4}_{-3}\times10^{-4}$Hz.
Note particularly the lack of a second break, about a decade below $\nu_{B}$,
which is characteristic of a low state GBH. 

\begin{figure}
\psfig{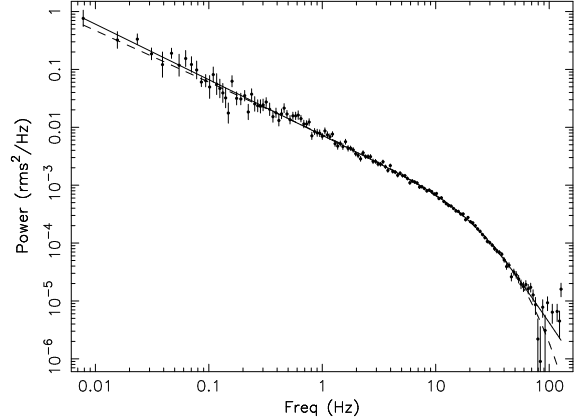}
\caption{\footnotesize PSD of Cyg~X-1 in the high state in the 2-13 keV range. 
The dashed line is a powerlaw with an exponential cut-off. The
solid line is a bending powerlaw, which
is a marginally better fit.  
Note the very strong similarity to the PSD of
NGC~451 (Fig.~\ref{fig:xtexmm012psd})}
\label{fig:cygpsd}
\end{figure}

In order to provide a good template for comparison we present,
in Fig~\ref{fig:cygpsd}, the PSD of the GBH Cyg X-1 in its high
state.  It is fitted very well by the same bending powerlaw model
which fits Fig~\ref{fig:xtexmm012psd}. We therefore conclude
that NGC~4051 is the first AGN confirmed to be in a `high' state. 

In Cyg X-1 (Fig~\ref{fig:cygpsd}) $\nu_{B}=22\pm1$Hz.
If $\nu_{B}$ scales linearly with mass then, assuming a black hole
mass of $10 M_{\odot}$ for Cyg X-1 \cite{her95},
we imply a black hole mass of $3^{+2}_{-1} \times 10^{5}
M_{\odot}$ in NGC~4051, which is consistent with the recently reported
reverberation value of $5^{+6}_{-3} \times 10^{5} M_{\odot}$ \cite{shemmer03}.
Hence NGC~4051 is emitting at $\sim30\% \rm~ L_{Edd}$.

\subsection{Variation of high frequency \xmm PSD with energy}
\label{sec:xmmpsd}

\begin{figure}
\psfig{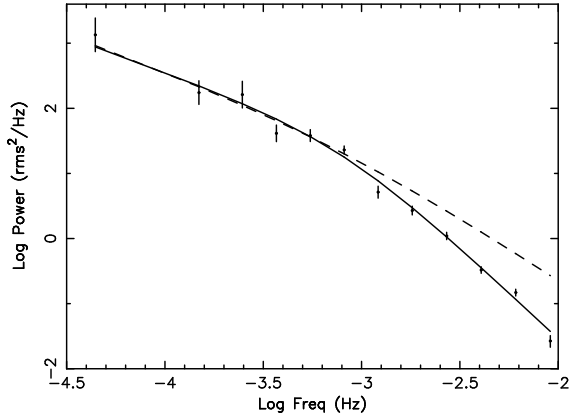}
\caption{\footnotesize PSD of NGC~4051 as measured by {\em XMM-Newton} 
in the 0.1-2 keV energy band. The solid line is the best fit with
$\alpha_{L}$ fixed at -1.1.  The Poisson noise level
has been subtracted from the PSD. The dashed line is the best 2-10 keV
fit with normalisation adjusted so that the PSDs agree at low
frequencies.}
\label{fig:0.1-2psd}
\end{figure}

We made PSDs of NGC~4051 in a number of different \xmm wavebands. All PSDs showed
a break at  $\nu_{B}\sim 10^{-3}$ Hz and had similar slopes
below $\nu_{B}$. As the PSD slope below $\nu_{B}$ in Cyg X-1 in the
high state is also independent of energy \cite{cui97b} we fixed 
$\alpha_{L}$ at -1.1, the best fit value from Fig~\ref{fig:xtexmm012psd}.
Further fitting confirmed the invariance of $\nu_{B}$ with energy but
showed a systematic flattening of $\alpha_{H}$ with increasing energy
(Fig~\ref{fig:0.1-2psd}).
Similar behaviour is seen in MCG-6-30-15 \cite{vfn03}.

Flattening of the PSD with increasing energy at high frequency is 
seen in Cyg X-1 in the low state \cite{nowak99}. In the generally 
accepted Compton scattering scenario for the production of X-rays 
the flattening is generally interpreted as arising from variations 
in the scattering corona as the increased scattering necessary to 
raise photons to higher energies would wash out variations from 
the seed photons. 

\section{THE RMS-FLUX RELATIONSHIP}

Uttley and \mch
\cite{uttley01} showed that the amplitude of
absolute rms variability in small segments of the X-ray lightcurves of
GBHs scaled remarkably linearly with the mean flux of the
segments. They also showed that the variability amplitude in {\it
RXTE} lightcurves of Seyfert galaxies increased with X-ray flux, but
could not tell
if the relationship was really linear.  Subsequently, better quality
lightcurves have provided stronger evidence of a linear rms-flux
relation in AGN \cite{edelson02,vfn03}. 

\begin{figure}
\psfig{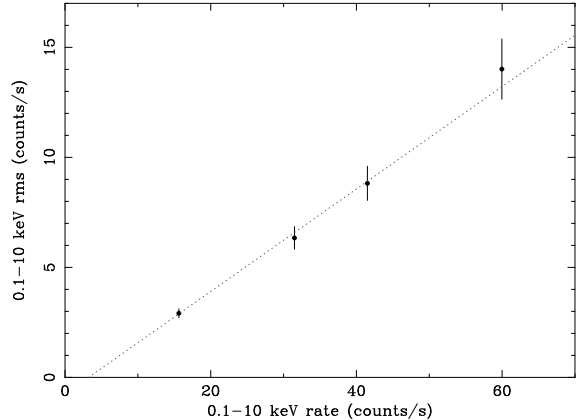}
\caption{\footnotesize rms-flux relationship for NGC~4051 derived from the \xmm
observations in the 0.1-10 keV band.}
\label{fig:rmsflux}
\end{figure}

We measured the mean flux and the noise-subtracted variance of 39
separate segments, of duration 2560-s, of the 5-s binned 0.1-10 keV
\xmm NGC~4051 lightcurve. We then averaged the variance in 4 
flux bins and took the square root to determine the rms
(Fig.~\ref{fig:rmsflux}). We note a strong linear rms-flux relationship, 
similar to that seen in GBHs, indicating a strong similarity in the
process driving the variability.

\section{TIME LAGS AND COHERENCE}

\begin{figure}[t]
\psfig{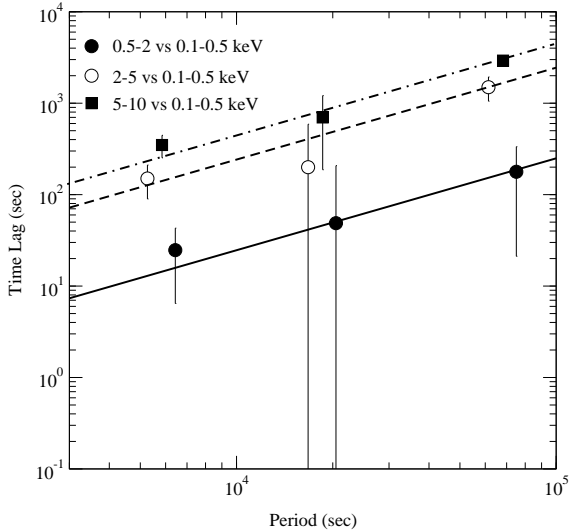}
\caption{\footnotesize Time lag vs. Fourier period for the cross spectrum of
the soft energy band ($0.1-0.5$ keV) vs. the $0.5-2, 2-5, $ and $5-10$
keV bands (filled, open circles and filled squares, respectively).
The hard bands are all delayed with respect to the soft (0.1-0.5~keV)
band. The solid, dashed, and dot-dashed lines show the best-fitting
power law models, assuming a slope of 1.  Note that the open and
filled circles have been displaced slightly in period so that
errorbars do not overlap.}
\label{fig:lags}
\end{figure}

Measurements of `time lags' and `coherence' between different 
wavebands provide a strong diagnostic of the geometry of the 
emission region (eg \cite{vn97,kotov01} and 
Section~\ref{sec:geom}). Here, for each waveband, we split the 
lightcurve into its Fourier components and measure the lag between 
variations in different bands, as a function of Fourier period. 
The results are shown in Fig~\ref{fig:lags}.
The hard band always lags the soft band and the lag 
increases with increasing Fourier period and with increasing 
energy separation of the bands (cf \cite{pap01}). 

\begin{figure}[t]
\psfig{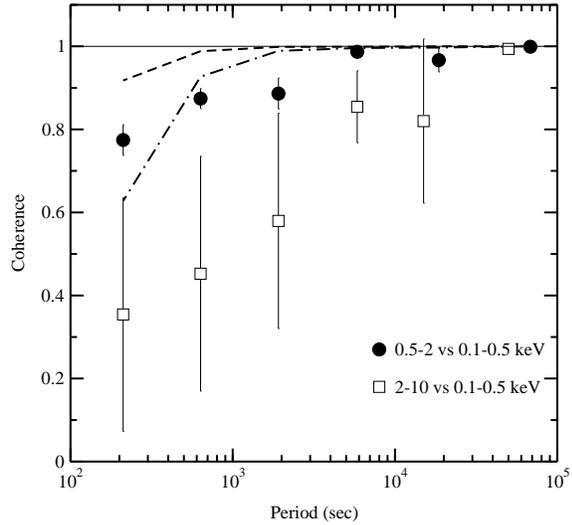}
\caption{\footnotesize Coherence function vs. Fourier period for the soft
energy band ($0.1-0.5$ keV) vs. the $0.5-2 $ and $2-10$ keV bands
(filled circles and open squares, respectively). The two open squares
at the longest Fourier period have been displaced in period to avoid
overlapping of errorbars. The dashed and dot-dashed lines show, for the
$0.1-0.5$ keV vs. $0.5-2 $ and $2-10$ keV bands respectively, the 95\%
confidence limit for spurious lack of coherence introduced by the
approximation in the numerical calculation. }
\label{fig:coherence}
\end{figure}

The coherence between two lightcurves, at a certain frequency, may be
interpreted as the correlation coefficient between the Fourier
components of the two lightcurves at that frequency \cite{p81,vn97}.
For NGC4051 we note (Fig~\ref{fig:coherence})
that, even taking account of the artificial decrease in coherence
introduced by numerical approximations, the coherence decreases with
increasing band separation and with increasing Fourier period.

\section{DISCUSSION}
\label{sec:disc}
\subsection{Geometry of the Emission Region}
\label{sec:geom}

The linear rms-flux relationship tells us that the amplitude of short
timescale variations responds to the amplitude of longer timescale
variations.  This response is explained naturally in the context of
the model of Lyubarskii \cite{lyub97} where a characteristic timescale is
associated with each radius of the accretion disc, with shorter
timescales arising closer to the black hole.  Variations propogate
inwards and modify the amplitude of shorter timescale variations. If
we also associate different photon energies with different
locations, with the bulk of the higher energies being produced closer
in, we can explain the observations of lags and coherence
mentioned earlier. 

Churazov et al. \cite{churazov01},
with enhancements by Kotov et al. \cite{kotov01}, have
built a qualitative model along these lines.  In the Kotov model
variations propogate inwards until they hit the X-ray emitting region,
whose emission they modulate. As a larger
fraction of the lower energy emission is produced further out in the
disc, where characteristic timescales are longer, we expect larger
lags with increasing energy separation of the bands and at longer
Fourier periods.  Regarding the coherence then, if the highest
energies are preferentially produced at the innermost radii, those
energies see the largest spectrum of variations, including those at
the shortest periods. However if the perturbations only propagate
inwards then the lower energies do not see the shortest period
variations and the coherence then decreases as we go to shorter periods and
greater separation of the bands.

\subsection{AGN black hole masses and break timescales}

We have gathered data from the literature (eg 
\cite{mark03,uttley02,pap02,vfn03,vf03})
and have plotted the timescale at which the PSD
breaks from a slope of $\sim-1$ to a slope of $\sim -2.5$, against the
best available value of the black hole mass.  This break might be the
only break in a high state PSD or it might be the higher frequency
break in a low state PSD. For no AGN but NGC~4051 can we be sure which break it
is. We also plot the break for Cyg X-1 in its high and low states. We
plot broad line AGN with filled circles and NLS1s with open
circles. The best fit line to all of the AGN together (not shown) does
not go through either state of Cyg X-1. 

The data are not sufficient to perform meaningful fits to subsets 
of the data but we can plot lines of linear scaling through both 
the high and low states of Cyg X-1. We note that the broad line 
AGN are consistent with linear scaling from Cyg X-1 in its low 
state but the NLS1s scale better to the high state. We therefore 
suggest that some parameter varies between the high and low states 
in such a way as to move the line of the scaling relationship to 
shorter timescales for the NLS1s/high state systems. The break 
timescale in NGC4051 (1250s) corresponds approximately to the 
viscous or thermal timescale at $\sim$few $R_{G}$ (cf 
\cite{treves88}). In GBHs the inner edge of the accretion disc 
moves inwards as the accretion rate increases (eg 
\cite{mcclintock03}). If the break timescale is therefore 
associated with the inner edge of the accretion disc, the 
parameter which moves the position of the mass-timescale scaling 
relationship might well be the accretion rate, $\dot{m}$.

\begin{figure}[t]
\psfig{figure=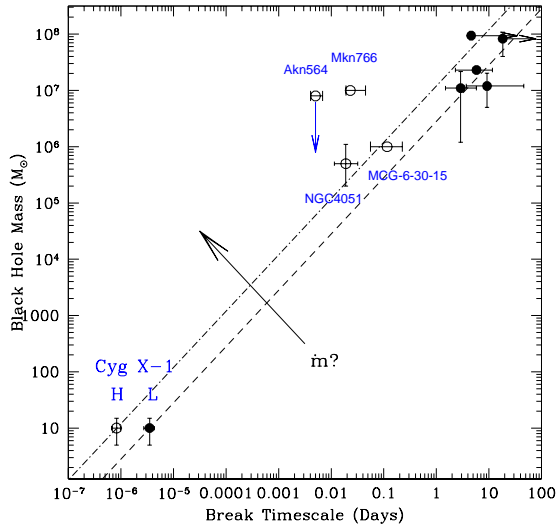,width=7.5cm,angle=0}
\caption{\footnotesize  PSD break timescales vs black hole mass. NLS1s are
shown as open circles and broad line Seyferts are shown as filled
circles.  The high (H) and low (L) states of Cyg~X-1 are
also plotted and lines (dot-dash and long-dash respectively) of slope
1.0 are drawn through those points.  The solid arrow labelled with
$\dot{m}$ indicates the way that the break timescale/mass line may
move with increasing accretion rate.}
\label{fig:bh}
\end{figure}

{}

\end{document}